\documentclass[11pt]{article}
\usepackage[cp866]{inputenc}

\makeatletter

\renewcommand{\@oddhead}{\hfil Alexander Shatskiy}

\input epsf

\begin{document}

\begin{center}
{\bf \Large  Is There a Relationship between the Density of
Primordial Black Holes in a Galaxy and the Rate of Cosmological Gamma-Ray Bursts?}\\
\end{center}

\begin{center}
Alexander Shatskiy *
\end{center}
*Astro Space Center, Lebedev Physical Institute, Russian Academy
of Sciences, Russia


DOI: 10.1134/S1063772906100015

$ \quad$  \\

{\bf \Large Abstract}\\
The rate of accretion of matter from a solar-type star onto a
primordial black hole (PBH) that passes through it is calculated.
The probability that a PBH is captured into an orbit around a star
in a galaxy is found. The mean lifetime of the PBH in such an
orbit and the rate of orbital captures of PBHs in the galaxy are
calculated. It is shown that this rate does not depend on the mass
of the PBH. This mechanism cannot make an appreciable contribution
to the rate of observed gamma-ray bursts. The density of PBHs in
the galaxy can reach a critical value - the density of the mass of
dark matter in the galaxy.

\section{INTRODUCTION}
There have been several important developments in astronomy over
the past two decades associated with the discovery of objects that
are candidate black holes. Black holes with galactic masses
(${10^6\div 10^9M_\odot}$), stellar masses (${\sim1M_\odot}$), and
intermediate masses (${\sim 10^3M_\odot}$) have all been
identified. Naturally, black holes with masses that are much less
than a solar mass can be detected only indirectly (such as via
gravitational microlensing). So-called primordial black holes
(PBHs) are in this class. They can be formed only in the early
stages of evolution of the Universe, since the natural evolution
of a star does not permit the formation of black holes with masses
less than a solar mass.

A black hole will unavoidably "devour" any matter in the immediate
vicinity of its gravitational radius (more precisely, within about
$3r_g$). Therefore, if a PBH is located inside a star, the life of
the star will be shortened by some time. After the passage of this
time, all the remaining matter of the star will unavoidably
collapse into the black hole.

One consequence of the collapse of the star into such a PBH should
be a powerful gamma-ray outburst, which is emitted by the remnants
of the stellar material. It is possible that such gamma-ray bursts
contribute to all the cosmological gamma-ray bursts that are
observed in the Universe on a continual basis.

Knowing the rate of accretion of matter onto a PBH and the
distribution of PBHs in space, we can estimate the rate of
associated gamma-ray bursts and compare it with the observed rate,
and also estimate the number of PBHs in a galaxy.

\section{RATE OF ACCRETION OF A STAR ONTO A PBH}

A model for accretion onto a black hole has been considered in
detail in \cite{Bondi}, \cite{Sh-T}, and we present here only the
main conclusions of this theory.

We consider the hydrodynamical theory of accretion onto a black
hole. The gravitational radius $r_g$ of the PBH with mass $m$ is
assumed to be much smaller than the scale on which the
self-gravitation of the star becomes stronger than the gravitation
of the PBH. We denote this scale $R_{sg}$; note that it should be
much smaller than the size of the star itself $R_\odot \approx
7\cdot 10^{10}$sm, (for a star of mass $M_\odot \approx 2 \cdot
10^{33}$g). We then have:
\begin{equation}
r_g \ll R_{sg} \ll R_\odot \, , \quad R_{sg} =R_\odot \left( {r_g
\over R_g} \right)^{1/3}  \, , \quad r_g = {2Gm \over c^2} \, ,
\quad R_g = {2GM_\odot \over c^2} \, . \label{2-1}
\end{equation}
The rate of accretion onto a PBH with mass $m$ is equal to (the
Bondi solution):
 \begin{equation}
\dot m = {4\pi \lambda (G m)^2 \rho_\odot \over c^3 \left( a^2 +
\beta^2_p \right)^{3/2} } \, . \label{2-2}
\end{equation}
Here, $\lambda$ is a constant of order unity, $ca$ is the sound
speed in the star (${c^2a^2 = dP/d\rho_\odot}$), $c \beta_p$ is
the speed with which the PBH moves through the star (at the
perigee of its orbit, $a<<\beta_p$), $\rho_\odot$ is the mean
density of matter in the star. Thus, during its passage through
the star, the PBH "digs a tunnel"\, in the stellar material with a
cross-sectional area
\begin{equation}
s={\lambda\pi r_g^2 \over \beta_p (a^2 + \beta_p^2)^{3/2}  }
\approx {\pi r_g^2 \over \beta^4_p} \, . \label{2-3}
\end{equation}
The associated change in the star's mass is
\begin{equation}
\Delta m \approx \rho_\odot s R_\odot \approx m\cdot {3 r_g R_g
\over 4 \beta^4_p R^2_\odot} \, . \label{2-4}
\end{equation}
During this accretion, the internal parameters of the star
(dependence of the temperature and density on radius) can be taken
during computations to be equal to the solar values from
\cite{sbor1}. Questions related to the stationarity of the
accretion are considered in Appendix 1.

\section{CAPTURE OF A PBH IN A STELLAR
ORBIT (ORBITAL CAPTURE)}

Moving in a galaxy with virial speeds of the order of
${c\beta_\infty \approx 300}$ km/s (${\beta_\infty \approx
10^{-3}}$), a PBH has some probability to collide with stars. The
cross section for such a collision $S$ is $\pi$ multiplied by the
square of the maximum impact parameter at infinity, for which the
PBH passes "within half" the radius of the star. The quantity $S$
is determined by equating half the stellar radius and the distance
from the star to the PBH at the perigee of its orbit for the
motion of the PBH along a hyperbolic trajectory:
\begin{equation}
S=\pi R^2_\odot \left( {1\over 4} + {R_g\over 2R_\odot
\beta^2_\infty} \right) \approx 2.4 \pi R^2_\odot \, . \label{7-1}
\end{equation}
Before the collision, the speed of the PBH is more than a third of
the cosmic velocity for such a star. The speed at infinity
$\beta_\infty$ is related to the speed just before the collision
(at perigee) by the formula
\begin{equation}
\beta^2_p = \beta^2_\infty + {R_g\over R_\odot} \,  , \quad \to
\quad \beta_p \approx 2.3\beta_\infty \, . \label{7-1-2}
\end{equation}
Since momentum is conserved in collisions involving the PBH, its
speed will decrease as its mass increases. Thus, its speed
immediately after a collision will be lower than it was just
before the collision. The collisions will continue until the PBH
is captured into an orbit by a star, i.e., until the speed of the
PBH turns out to be less than a third of the cosmic speed after a
collision: ${\beta_3 = \sqrt{ R_g / R_\odot } \approx
2.1\beta_\infty}$.

Thus, the total change in the speed of the PBH required for
orbital capture should be
\begin{equation}
\Delta\beta = \beta_p - \beta_3 \approx 0.2\beta_\infty \, .
\label{7-1-3}
\end{equation}
In accordance with the conservation of momentum and expression
(\ref{2-4}), the change in speed during the collision process is
determined by the formula
\begin{equation}
\Delta\beta_1 \approx \beta_p {\Delta m\over m} \approx {3 r_g R_g
\over 4 \beta^3_p R^2_\odot} \approx {r_g R_g \over 16 R^2_\odot
\beta^3_\infty} \, . \label{7-1-4}
\end{equation}
Accordingly, the number of collisions required for orbital capture
is
\begin{equation}
k = {\Delta\beta \over \Delta\beta_1} \approx 3.2{\beta^4_\infty
R^2_\odot \over R^2_g}\cdot {R_g\over r_g} \approx {0.2 M_\odot
\over m} \, . \label{7-1-5}
\end{equation}
We can see that, in order for an orbital capture to occur as a
result of the first collision, the mass of the PBH before the
collisionmust ${m\approx 0.2 M_\odot}$. In this case, the mass $m$
of the PBH will approximately double as a result of the collision.

\section{RATE OF ORBITAL CAPTURES OF PBH
IN A GALAXY}

We can determine the mean time $T_1$ between collisions using the
formula ${S \beta_\infty c T_1 =V_0}$, where ${V_0 = R^3_0}$
 is the mean volume per star in the galaxy
(${R_0 \approx 10 \makebox{ light years }\approx 10^{19}}$sm):
\begin{equation}
T_1 = {V_0 \over S \beta_\infty c} \approx 3.8\cdot
10^{23}\makebox{years} \, . \label{7-2}
\end{equation}
Thus, the time for the mean free path of a PBH (until its orbital
capture) is ${\tau = k\cdot T_1}$. If the total number of
uncaptured PBHs in the galaxy is $N_{bh}$, some number of them
($N_t$) will be captured by stars over a time $t$:
\begin{equation}
N_t = N_{bh}\cdot \left[ 1 - \exp (-t/\tau) \right] \approx
N_{bh}\cdot t /\tau \, . \label{7-2}
\end{equation}
The number of PBHs in the galaxy is given by the relation
${N_{bh}=\kappa M_{tot}/m}$, where ${M_{tot} \approx 10^{46}}$g is
the total mass of the galaxy (assuming it is similar to the
MilkyWay) and $\kappa$ is a coefficient determining the mass
fraction of PBHs in the galaxy (${0 < \kappa < 1}$). We can see
that (\ref{7-2}) does not depend on the mass of the PBH. The rate
of orbital captures of PBHs in the galaxy is then
\begin{equation}
{dN_t\over dt}\equiv \dot N_t \approx {2.4\kappa c R_g \over V_0
\beta^3_\infty}\cdot {2G M_{tot} \over c^2} \approx \kappa\cdot
{1\over 10^{6}\makebox{years}} \, . \label{7-3}
\end{equation}
Thus, the rate of orbital capture can reach values of the order of
one capture every million years in a MilkyWay-type galaxy.

\section{LIFETIME OF A PBH IN ORBIT
AROUND A STAR}

If the lifetime of a PBH in orbit is less than the lifetime of the
star (${\sim 10^9}$yrs), PBHs will not be accumulated in the
galaxy.

The lifetime of a PBH in orbit is determined by the increase of
its mass by a factor of two due to accretion. The corresponding
number of revolutions $n$ of the PBH around the star is given by
the relation
\begin{equation}
n = {m\over \Delta m} \approx {37 \beta^4_\infty R^2_\odot \over
R_g r_g} \approx {2 M_\odot \over m}\, . \label{3-4}
\end{equation}
The period of revolution about the star $\tau_0$  is determined
from the assumption that the major axis of the orbit is ${\sim 0.5
R_0}$ (half the mean distance between stars):
\begin{equation}
\tau_0 \approx \pi R_0^{3/2}/(c\sqrt{R_g}) \approx 1.7 \cdot 10^8
\makebox{yrs } \, . \label{3-4-2}
\end{equation}
We thus obtain for the lifetime in the orbit
\begin{equation}
T_0 = n \tau_0 \approx {117 \beta^4_\infty R^2_\odot \over R_g
c}\cdot \left( {R_g\over r_g} \right) \cdot \left( {R_0\over R_g}
\right)^{3/2} \approx 3.5 \cdot {M_\odot\over m} \cdot
10^8\makebox{years} \, . \label{3-5}
\end{equation}
The masses of PBHs can be constrained from above by data available
from gravitational microlensing observations: ${m<0.1M_\odot}$
(see~\cite{lens1}). Therefore, in order for the time $T_0$ to be
less than the lifetime of the Universe (${\sim 15\cdot 10^9}$yrs),
the mass of a PBH must lie in the very narrow range of $0.03
M_\odot$ to $0.1 M_\odot$. In this case, gamma-ray bursts induced
by the collapse of a star into a PBH will occur from time to time
in a galaxy.

However, even in this case, this mechanism will not make an
appreciable contribution to the observed gamma-ray bursts in a
galaxy, since the rate of such events (\ref{7-3}) is much less
than the mean rate of gamma-ray bursts in a galaxy ${\sim 0.001
\div 0.01}$ yrs${}^{-1}$.

It follows from the above analysis that PBHs can comprise any mass
fraction of the dark matter in a galaxy.

\section{Appendixes}

\subsection{TIME FOR ESTABLISHING
A QUASI-STATIONARY ACCRETION REGIME AS A PBH IMPACTS A STAR}

The relative speed between the PBH and the particles in the
stellar medium changes as a PBH moves through a star uniformly and
linearly. Consequently, there are relative accelerations, which
correspond to inertial forces. If we denote the coordinate of the
PBH along its trajectory ${y=c\beta_p t}$ and the minimum distance
between the particles in the stellar medium and the PBH trajectory
$x$, this acceleration $g$ can be expressed in terms of $x$ and
$y$ as follows:
\begin{equation}
|g| = {(c \beta_p x)^2 \over (x^2 + y^2)^{3/2} } \, . \label{a1-1}
\end{equation}
Here, it is convenient to assign the initial value of $y$ to be at
a point that is closest to a particle of the medium. The condition
for capture of a particle in the medium by the passing PBH is
given by the inequality
\begin{equation}
|g| < {Gm\over x^2 + y^2} \, . \label{a1-2}
\end{equation}
This corresponds to the dominance of the gravitational force over
the inertial force, and must be satisfied at any moment in time.
We thus obtain
\begin{equation}
x < x_0 = {r_g \over 2\beta^2_p } \, . \label{a1-3}
\end{equation}
This qualitative expression approximately corresponds to the cross
section $s=\pi x_0^2$ [see~(\ref{2-3})].

The relaxation time for the stellar material during a collision
with a PBH is determined (to order of magnitude) by the free-fall
time for a particle to fall from a distance $x_0$ onto the PBH:
\begin{equation}
t_0 = {r_g \over 3\sqrt{2} c\beta^3_p }\, . \label{a1-4}
\end{equation}
We have for a mass of ${m<0.1M_\odot}$ the time ${t_0 < \sim
10}$s. Thus, this time is much shorter than the time for the
passage of the PBH through the internal layers of the star:
${\left[ \sim R_\odot /(c\beta_p)\right] \sim 1000}$s.
Consequently, the accretion process very rapidly becomes
quasi-stationary.

\subsection{STATISTICS OF THE MASS AND SPEED
DISTRIBUTIONS OF PBHS IN THE UNIVERSE}

The nature of PBHs in the early Universe (if indeed they exist) is
not clear, and there are also no precise models describing their
statistical distribution. Moreover, we cannot consider the early
Universe to be in some kind of thermodynamic equilibrium.
Therefore, the arguments presented below do not pretend to be the
only valid description of the statistics of PBHs, and this has led
us to present them in a separate appendix.

Let us consider a system of a larger number of PBHs in a spatial
volume $V$. We will consider this volume to be sufficiently large
so that we can apply a statistical analysis to the PBHs. Their
masses are bounded from below by the relation
${m_{min}=E_{min}/c^2}$ because of the condition for the Hawking
evaporation of black holes.

Let us assume that the PBHs have a canonical Gibbs
distribution~\cite{LL5}. Thus, the probability density for the
existence of a PBH with a mass m and a spatial velocity $v$ is
\begin{equation}
w=c_1 \exp \left\{ -{y \alpha \over \sqrt{1-\beta^2}}\right\}\, ,
\label{1-1}
\end{equation}
where ${y \equiv mc^2/E_{min}}$ is the dimensionless rest energy
of a PBH, ${\alpha \equiv E_{min}/T}$ is the dimensionless inverse
temperature of the system, ${\beta =v/c}$ the speed of the PBHs in
units of the speed of light, and $c_1$ is a normalization
constant.

The normalization condition is
\begin{equation}
\int\limits_1^\infty dy \int\limits_0^1 4\pi\beta^2\, d\beta \, w
\int\limits_0^{V} d^3 r =1. \label{1-2}
\end{equation}
We now introduce a number of functions that depend on  $\alpha$:
\begin{equation}
I_n (\alpha )=\int\limits_0^1 4\pi\beta^2\, d\beta \, \left[
\sqrt{1-\beta^2} \right]^n \exp \left\{ -{\alpha \over
\sqrt{1-\beta^2}}\right\}\, ,\quad I'_n(\alpha ) =-I_{n-1}(\alpha
) \, . \label{1-3}
\end{equation}
The condition (\ref{1-2}) then takes the form
\begin{equation}
c_1V I_1 =\alpha\, . \label{1-4}
\end{equation}
We now introduce the mean energy of the PBH:
\begin{equation}
U=\left< {y\over\sqrt{1-\beta^2}} \right> = \int\limits_1^\infty
dy \int\limits_0^1 4\pi\beta^2\, d\beta \,
{y\over\sqrt{1-\beta^2}} c_1V \exp \left\{ -{\alpha y \over
\sqrt{1-\beta^2}}\right\} \, . \label{1-5}
\end{equation}
After integrating over $y$, this expression can be written in the
form
\begin{equation}
U\alpha^2 =c_1V \left[ I_0 \alpha + I_1\right] \, . \label{1-6}
\end{equation}
The energy and entropy of any system are additive quantities,
equal to the sums of the contributions from their individual
parts. Black holes should not be an exception to this behavior.
Therefore, the total energy and entropy of the system will have
the forms
\begin{equation}
U_{tot}= \int\limits_1^\infty dy \int\limits_0^1 4\pi\beta^2\,
d\beta \, N{y\over\sqrt{1-\beta^2}} c_1V \exp \left\{ -{\alpha y
\over \sqrt{1-\beta^2}}\right\} = NU\, . \label{1-7}
\end{equation}
\begin{equation}
S_{tot}= \int\limits_1^\infty dy \int\limits_0^1 4\pi\beta^2\,
d\beta \, NS_0 c_1V \exp \left\{ -{\alpha y \over
\sqrt{1-\beta^2}}\right\} = N \left< S_0 \right> \, . \label{1-8}
\end{equation}
where $N$ is the total number of PBHs in the system, $S_0=c_2 y^2$
is the entropy of a PBH with mass $m$, and $c_2$ is a
dimensionless coefficient~\cite{membr}:
\begin{equation}
c_2= {4\pi G m^2_{min}\over \hbar c} \gg 1\quad \makebox  { at  }
m_{min}\gg 10^{-5}g. \label{1-8-1}
\end{equation}
Thus, ${S_{tot}=\left < S_0 \right> U_{tot}/U}$. Using the
thermodynamic expression for the temperature, ${1/T =
dS_{tot}/dE_{tot}}$ (see~\cite{LL5}), we obtain
\begin{equation}
\alpha = {dS_{tot}\over dU_{tot}}= \int\limits_1^\infty dy
\int\limits_0^1 4\pi\beta^2\, d\beta \, {c_1V c_2\over U} y^2 \exp
\left\{ -{\alpha y \over \sqrt{1-\beta^2}}\right\} \, .
\label{1-9}
\end{equation}
Integrating over $y$, we obtain
\begin{equation}
U\alpha^4 =c_2c_1V \left[ 2I_3 +2\alpha I_2 +\alpha^2 I_1 \right]
\, . \label{1-10}
\end{equation}
Equations (\ref{1-4}), (\ref{1-6}) and (\ref{1-10}) can be used to
find $\alpha$ and $U$, e.g., by multiplying (\ref{1-6}) by
$\alpha^2$ and substituting this into (\ref{1-10}). We
differentiate the resulting expression with respect to $\alpha$ to
obtain
\begin{equation}
\alpha^2 I_{-1} -\alpha (c_2 +2)I_0 -2I_1 =0 \label{1-11}
\end{equation}
This is a transcendental equation in $\alpha$. However, if $c_2
\gg 1$, then $\alpha \gg 1$. We will use this fact and multiply
the function $I_n$ by the factor ${\exp\left\{ \alpha
\right\}/(4\pi )}$. Denote the new functions $\tilde I_n$:
\begin{equation}
\tilde I_n (\alpha )=\int\limits_0^1 \beta^2\, d\beta \, \left[
\sqrt{1-\beta^2} \right]^n \exp \left\{ \alpha \left( 1-{1\over
\sqrt{1-\beta^2}}\right) \right\} \approx \exp \{ -\alpha\beta_0^2
/2\} \cdot \beta_0^3 \cdot\left( 1 - {\beta_0^2 n\over 2}\right)
\, . \label{1-12}
\end{equation}
In $\tilde I_n$, the integral is already rapidly cut off by the
exponential when ${\beta_0^2\approx 2\kappa /\alpha \ll 1}$, where
$\kappa$ is a coefficient of order 10. Thus, the functions
obtained can be expanded in a series in $\beta_0$, keeping only
the first term. The function $\tilde I_n$ preserves the form of
(\ref{1-11}). Substituting the asymptotic (\ref{1-12}) into
(\ref{1-11}) yields
\begin{equation}
\alpha^2 - \alpha (c_2 + 2 - \kappa )  -2 + 2\kappa/\alpha =0\, .
\label{1-13}
\end{equation}
We find in the main approximation
\begin{equation}
\alpha\approx c_2 \, ,\quad U\approx 1 +5/(4\alpha) \approx 1 \, .
\label{1-14}
\end{equation}
Hence, the temperature of the Gibbs distribution of PBHs is of
order the Hawking temperature ($T_{_H}$) for $m_{min}$:
\begin{equation}
T \equiv {E_{min}\over \alpha} \approx {E_{min}\over c_2}= {\hbar
c^3\over 4\pi Gm_{min}}= 2T_{_H}= 2,4\cdot\left( {10^{26}g\over
m_{min}}\right) \, K{}^\circ \label{1-15}
\end{equation}
It follows that the bulk of PBHs have very narrow ranges of their
mass (from $m_{min}$ to ${m_{min}(1+1/\alpha )}$) and velocity:
$$
\makebox{from 0 to } \beta_0 \sim {10^{-4} g \over m_{min} } \ll
\beta_\infty \ll 1 .
$$
The probability density for the PBHs falls off exponentially
outside these intervals. With such a low value for $\beta_0$, its
contribution to the motion of the PBH can be neglected compared to
the virial speeds in the galaxy, which are determined by the
gravitation of all its bodies.

Thus, to a high degree of accuracy, this model for PBHs represents
an ideal gas with particles of mass $m_{min}$ and velocities of
the order of the virial velocity in the galaxy.

\bigskip\hrule\bigskip

{\bf  \Large ACKNOWLEDGMENTS }\\

The author thanks N.S. Kardashev, V.N. Lukash, B.V. Komberg, D.A.
Kompaneits, and other participants in seminars at which this work
was discussed, refined, and augmented.

The work was supported by the Program "Non-stationary Phenomena in
Astronomy, 2005"\, (no. NSh-1653.2003.2) and the Russian
Foundation for Basic Research (project nos. 05-02-17377,
05-02-16987-a).

\end{document}